\documentstyle[12pt]{article}
\newcounter{mycount}

\newcommand{\be}{\begin{eqnarray}}
\newcommand{\ee}{\end{eqnarray}}
\newcommand{\bfl}{\begin{flushleft}}
\newcommand{\efl}{\end{flushleft}}
\newcommand\ie{{\it i.e. }}

\newcommand\etal{{\it et.al.}}

\newcommand\half{\frac 1 2 }

\newcommand\noi{\noindent}

\begin{document}

\centerline{\Large\bf Baryonic and Gluonic Correlators in Hot QCD}
\vspace* {-35 mm}
\begin{flushright} USITP-94-1 \\
SUNY-NTG-93-12 \\
January 1994
\end{flushright}
\vskip 0.9in
\centerline{{\bf T.H. Hansson$^{1}$, M. Sporre$^2$ and
  I. Zahed$^3$} }
\vskip 20mm\noi
\centerline{\bf ABSTRACT}
\vskip 5mm
We extend our earlier work on static color singlet correlators in high T QCD
(DeTar correlators) to baryonic and gluonic sources, and estimate the
corresponding screening masses using the dimensionally reduced theory.
We discuss spin and
polarization dependence of meson and baryon masses in the $T \rightarrow
\infty$ limit,  and possible nonperturbative effects at non-asymptotic
temperatures.

\vfil\noi
$^1$
Institute of Theoretical Physics, University of Stockholm \\
Fysikum, Box 6730, S-113 85 Stockholm, Sweden \\
email: Hansson@vana.physto.se
\vskip 2mm\noi
$^2$ NORDITA \\
Blegdamsvej 17 \\
DK-2100 Copenhagen O, Denmark \\
email: Sporre@NORDITA.dk
\vskip 2mm\noi
\noi$^3$Nuclear theory group, Department of physics, SUNY at Stony Brook \\
Stony Brook, New York, 11794, USA \\
email: Zahed@sbnuc1.physics.edu
\vskip 3mm \noindent
${^1}$Supported by the Swedish Natural Science Research Council. \\
${^3}$Supported in part by the Department of Energy under Grant \\
 No. DE-FG02-88ER40388.

\eject

\bibliographystyle{nphys}

\noi{\large\bf 1. Introduction }
\renewcommand{\theequation}{1.\arabic{equation}}
\setcounter{equation}{0}

Recent lattice measurements of correlation functions of the type
\be
C_{\alpha} ( x_3 )=
\langle \int_0^{\beta}d\tau\int d^2 r\, J_{\alpha} (\vec r,x_3, \tau )\,\,
\int_0^{\beta}\,d\tau' J_{\alpha} (\vec 0,0, \tau' )\rangle
\ee
(where $\beta = 1/T$, $\vec r = (x_1,x_2)$ and $J_\alpha$
a color-singlet operator)\footnote{
For baryonic sources the average over imaginary time, $\tau$, is weightened
by $\cos\omega_0\tau$ where $\omega_0$ is the lowest Matsubara frequency.}
show that  the hadronic screening lengths above the QCD
finite temperature (chiral) phase transition point fall into
chiral multiplets. The screening masses are approximately $2\pi T$ for
mesons and $3\pi T$ for baryons, with the notable exceptions being
the pion and its scalar partner, which are
considerably lower \cite{gava1}.
It has also been known for some time that bulk thermodynamic quantities, like
energy and entropy density, are roughly as expected from a gas
of non-interacting quarks and gluons \cite{kars1}.
The exception here is the pressure,
which is considerably lower than expected from the free gas picture. Finally,
the fermionic susceptibility, $\partial^2 \ln Z / \partial m^2 $ varies rapidly
across the transition region from being small in the hadron phase, to
being large  in the quark phase \cite{gott1}.
All these results are suggestive of a high temperature phase of weakly
interacting (Debye screened) quarks and gluons.

On the other hand, the Coulomb gauge lattice calculations
of the spatial structure of the mesonic correlators via
\be
C_{\alpha} (\vec r, x_3 )=
\langle \int_0^{\beta}d\tau\int d^2 r_1\,d^2 r_2\,
{\psi}^{\dagger} (\vec r_1,0, \tau ) \Gamma_\alpha\psi (\vec r_2,0, \tau )
\nonumber \\
\int_0^{\beta}d\tau'\,  \psi^{\dagger} (\vec R,x_3, \tau' )
\Gamma_\alpha\psi (\vec R-\vec r,x_3, \tau' )  \rangle
\ee
show strong evidence for correlations in the
transverse direction \cite{bern1}.
The presence of such correlations is not unexpected given
that the the spatial Wilson loops obey an area law at all
temperatures \cite{mano1},\cite{kark1},\cite{fing1}.

What causes the correlations in the spatial
directions, and to what extent they are important
for our understanding of the high temperature phase, is not clear.
What is clear, however, is that any description of the high temperature
phase has to account for the above results.
In a recent paper \cite{hans5}, hereafter referred to as I,
two of us suggested that at very high temperature, the screening
lengths and ''wave functions'' discussed above, can be understood from an
analysis of the static part of the gluon field together with the lowest energy
quark modes. This ''dimensionally reduced'' QCD is a YM-Higgs model with heavy
(\ie $M=\pi T$) quarks, which is believed to be confining.\footnote{
This is based on the already mentioned fact that space-like Wilson loops obey
an area law at all temperatures. Recent work by K{\"a}rkk{\"a}inen \etal
\cite{kark1} shows that
the string tension $\sigma (T)\sim T^2$ at high temperature.}
The screening masses and wave functions correspond to masses and wave functions
of heavy quark states in the dimensionally reduced theory, and can be
calculated as Coulomb bound states using usual charmonium-type methods.
The masses, $m_\alpha$, in the mesonic
correlators are thus naturally $m_\alpha\approx 2M = 2\pi T$, and the wave
functions show
exponential fall-off at large distances, both in qualitative agreement with the
numerical calculations. It has also been shown, that in the case of 2+1
dimensional
QCD at high $T$, the Coulomb bound state picture leads to the correct
description of the quark susceptibilities at
asymptotic temperatures \cite{prak1},\cite{zahe1}.

In this paper we extend our analysis in I to
(color singlet) baryonic and gluonic sources and give a more detailed
discussion
of the spin and polarization dependence of the screening masses.
For the baryons the calculations are very similar to the mesonic case and are
on the same level of rigour. In the case of gluonic operators, our methods
are more questionable since the size of the resulting
bound state is larger than the Debye screening length, and we cannot rule out
the possibility that the infrared divergences in the magnetic sector will
change our predictions considerably.
In section 2 we recapitulate the arguments that lead to the dimensional
reduction scheme using a formalism that is somewhat different from the
one used in I. We also discuss the polarization dependence of
the various mesonic correlators.
Our results for baryonic and gluonic correlators can be found in
section 3 and 4 respectively. In the latter we also consider correlations
between Wilson lines.
In section 5, we discuss non-asymptotic effects that might
explain the difference between the lattice data and our asymptotic
prediction for the fine-structure of the screening mass spectrum. Section 6,
finally, contains some summary comments and discussion. Some details about
the calculations can be found in Appendix, A, B an C.

\vskip 3mm \noi
{\large\bf 2. Dimensional Reduction and DeTar Correlators }
\renewcommand{\theequation}{2.\arabic{equation}}
\setcounter{equation}{0}
\newcommand \ega {\hat\gamma}
\newcommand \hpsi {\hat\psi}
\newcommand \hpsid {\hat\psi^\dagger}
\newcommand \hphi {\hat\phi}
\newcommand \bphi {\overline{\phi}}
\newcommand \hphid {\hat\phi^\dagger}
\newcommand \udag {U^\dagger}
\newcommand \vdag {V^\dagger}
\newcommand \dagg {\dagger}

We now briefly summarize the dimensional reduction scheme
for the fermionic part of the high T QCD lagrangian. The starting point is the
Euclidian action for free fermions,
\be
{\cal L}_4 = i \hpsi^{\dag} (-\ega_\mu \hat D_\mu  + m) \hpsi \ \ \ \ \ ,
\ee
where $\hat D_\mu = \hat\partial_\mu -ig\hat A_\mu$, and where the (Hermitian)
Euclidian gamma matrices satisfy
\be
\{ \ega_\mu , \ega_\nu\} = 2\delta_{\mu\nu}
\ee
Now, retain only the modes with energy $\omega =\pm\pi/\beta = \pm \pi T$ and
define ${\hphi_\pm}$ by
\be
\sqrt\beta \hpsi &=&V U_+  e^{i\pi T\hat x_4} \hphi_+ +V U_-
e^{-i\pi T\hat x_4} \hphi_- \nonumber \\
\sqrt\beta\hpsi^\dagg &=& \hphi_+^\dagg U_+ \vdag e^{-i\pi T\hat  x_4} +
  \hphi_-^\dagg U_-\vdag e^{i\pi T\hat x_4 }
\ee
where $ U_\pm = e^{\mp i{\frac \vartheta 2}\ega^3}$ with $\cot  \vartheta =
m/\pi T$,
and where $V=e^{i\hat\sigma_{43}\pi/4}$ with $\hat\sigma_{\mu\nu} = \frac i 2
[\ega_\mu, \ega_\nu]$, rotates
an angle $\pi/2$ in the ($\mu\nu$) plane.
In particular
$\vdag \ega_3 V  = \ega_4$ and $\vdag \ega_4 V =-  \ega_3$.
Expressed in these variables the dimensionally reduced Lagrangian
becomes
\be
{\cal L}_3 &=&  \sum_{a=\pm}
-i{\hphi^{\dagg}_a} [ \ega_4 (\hat\partial_3 - ig\hat A_3) +
 \ega_i (\hat\partial_i - ig\hat A_i) - M   \\
  &+& ig\hat A_4 (\cos\vartheta \ega_3 - ai\sin\vartheta )] \hphi_a
\nonumber
\ee
where $M^2 = (\pi T)^2 + m^2 $.
The last step is now to rotate to a fictitious 2+1 dimensional Minkowski space
(from now on greek indices run from 0 to 2 and roman from 1 to 2)
by  $\hphi =  \phi$, $\hphi^\dagg =i  \bphi$,
$(\hat x_i,\hat x_3) = (x^i,ix^0)$,
$(\ega_i,\ega_3, \ega_4) = (-i\gamma_i, -i\gamma_3, -\gamma_0)$ and
$(\hat A_i, \hat A_3, \hat A_4) = \sqrt T (A_i, -iA_0, H)$,
where $H$ is the Higgs field,  $g_3 = \sqrt{T} g$ the 3d gauge coupling
constant and where we use the metric $(+--)$ in Minkowski space.
The resulting Lagrangian reads\footnote
{In I we used a different reduction scheme based on two-spinors.
The lagrangian given there was not correct due to a subtlety in one of the
transformations.  However,
the conclusions of the paper are not affected.
We give the correct two-spinor lagrangian in Appendix A.}
\be
{\cal L}_{2+1} = \sum_{a=\pm}
\bphi_a \left[i\gamma^\mu (\partial_\mu - ig_3A_\mu) - M + g_3(a \sin \vartheta
 - \cos\vartheta \gamma^3 ) H  \right] \phi_a \ \ \ \ \ .
\ee
In the dimensionally reduced theory the quarks are heavy with mass
$M$, the adjoint Higgs scalar has the electric mass
$m_E = \sqrt{4/3} gT$ (for two light fermions)
and the gluons are expected to acquire a magnetic mass $\sim g^2T$.
The quark and Higgs masses are easily understood in perturbation theory as a
consequence of the absence of fermionic zero modes and Debye screening
respectively. In
contrast, the magnetic mass is not calculable in perturbation theory and the
predictions made on the basis of the dimensionally reduced theory are only
reliable if they are insensitive to distance scales $\ge 1/g^2T$. A more
detailed discussion of these points can be found in I.

Note that our conventions are slightly unusual in that we use 4
dimensional matrices to represent the 2+1 dimensional Clifford algebra.
In Appendix A we give a formulation in terms of two different two-spinors
that however
have non-diagonal mass terms. For practical calculation the
Lagrangian (2.5) is more useful.

For $m=0$ the original Lagrangian (2.1) is invariant under the chiral
transformation
$\hpsi \rightarrow e^{i\alpha\hat\gamma_5}\hpsi$,
$\hpsi^\dagg \rightarrow \hpsi^\dagg  e^{i\alpha\hat\gamma_5}$
where $\ega_5 = \gamma_5 = -\ega_1\ega_2\ega_3\ega_4$. After the variable
change, this corresponds to the Lagrangian (2.5) being invariant under
$\phi_\pm \rightarrow e^{\pm i\alpha \gamma_3\gamma_5}\phi_\pm$.
Also note that in spite of the mass term ${\cal L}_{2+1}$ must be invariant
under parity transformations since the original ${\cal L}_4$ is. It is easy to
verify that the relevant transformation is $x_1\rightarrow -x_1$, $\phi_\pm
\rightarrow -i\gamma_5\gamma_1 \phi_\pm$.

Under the variable change (2.3) the expressions for the sources are also
changed
and we give a translation table for the most important currents in the
chiral limit $m=0$. Others are easily arrived at using (2.3):
\be
\hpsid \hpsi        &\rightarrow& \mp i\bphi_\pm\gamma_3\phi_\pm \nonumber \\
\hpsid \ega_5 \hpsi &\rightarrow& i \bphi_\pm\gamma_5\phi_\pm \nonumber \\
\hpsid \ega_4 \hpsi &\rightarrow& \mp \bphi_\pm\phi_\pm   \\
\hpsid \ega_3 \hpsi &\rightarrow& -i \bphi_\pm\gamma_0\phi_\pm \nonumber \\
\hpsid \ega_i \hpsi &\rightarrow&  \bphi_\pm\gamma_i\phi_\pm \ \ \ \ \  .
 \nonumber
\ee

As explained in I, the screening lengths and wave functions for mesonic and
baryonic operators, can be identified
with the masses and wave functions of the non-relativistic bound states
corresponding to (2.5). The relevant non-relativistic Hamiltonian is
obtained by
Breit reduction. Keeping only the leading Coulomb interaction we have
\be
H= \sum_i \,\frac {\vec{p}^2_i}{2M}\, +
    \sum_{i<j} \,\,\frac{e^2}{2\pi}\left( \frac 12 +
                     {\rm ln }(M|r_i -r_j|)\right)
\ee
where $e^2=g_{3}^2 C_F$ (quark-antiquark) and $e^2=g_{3}^2 C_F/2$
 (quark-quark) with $C_F$ the Casimir operator in the fundamental
representation. To arrive at this expression one has to cancel the infrared
singularities between self-energy and exchange contributions as explained in I.

We now give a formula for the spin-spin interaction in non-relativistic bound
states by performing a Breit reduction of the Hamiltonian corresponding to
(2.5). The easiest way to proceed is to notice that the expressions for the
currents are exactly as in the standard (3+1) dimensional
case, except that the momenta in the 3-direction are identically zero. Since,
in the nonrelativistic approximation, the spin-dependent part of the current
is $\chi^\dagger (\vec p') i\vec\sigma\times(\vec p' - \vec p)\chi(\vec p)$,
the Pauli interaction only depend
on the spin component in the 3-direction. A simple calculation yields,
\be
H_{ss} = \sum_{i<j} \,\, \frac{e^2}{4 M^2} (1 + 4 S_3^i S_3^j) \,\,
                                        \delta ^2 (\vec{r}_i - \vec{r}_j)
\ee
$H_{ss}$ is subleading in the temperature and, as usual, it will only be
considered as a perturbation. In the case of mesons (2.8) simplifies to
\be
H_{ss} = \frac{e^2}{2M^2} \,\,S_3^2  \,\,
           \delta^2(\vec r_1 - \vec r_2) \ \ \ \ \ ,
\ee
where $S_3$ is the total spin of the meson in the 3-direction.
As mentioned earlier, this expression differs from the one derived in I.
In the final result there is, however, only
one difference, namely
that the different polarization components of the rho
acquire different screening masses. Of the four components of the
interpolating massive
vector field $\rho_\mu = \hpsi \ega_\mu\hpsi$, the components 1, 2 and 4
are measured on the lattice \cite{deta2}.\footnote{
In \cite{deta2} $\rho_i$ and $\rho_4$ are referred to as $vt1$ and $vt0$
respectively, where the 0 and 1 denote the ''helicity'' which is the component
of the angular momentum about the 3-direction. We do not agree with this
helicity
assignment for $\rho_4$, as discussed in the text.}
Using the translation table (2.6) we see that these correspond to
$\bphi \gamma_i\phi$ and $\bphi \phi$. The first has spin-projection $\pm
1$, while the second is not an eigenstate of $S^2_3$, but a mixture of
1 and 0. The component $\rho_3$, which has not been measured on the lattice,
corresponds to the current $\bphi\gamma_0\phi$, and is a  $S^2_3=0$
eigenstate.

Using (2.9) we predict $m_\pi\simeq 2\pi T < m_{\rho_4}<m_{\rho_i}$ and
\footnote{
Similar results have also been obtained by Brown \etal
 using somewhat different methods.\cite{brow2}}
\be
m_{\rho_i} - m_\pi = \frac {e^2} {M^2} |\Psi(\vec 0 )|^2
\ee
The lattice measurements at temperatures not much above $T_c$ do not exhibit
this pattern but rather $\frac 3 2 m_\pi \simeq m_{\rho_i}
\simeq 2\pi T < m_{\rho_4}$.
One should remember, however, that our predictions are good at asymptotically
high temperatures, while the lattice calculations are performed not far
above the
transition temperature. The consequences of this will be discussed further in
section 6.

At this point it is also pertinent to make a few comments on the flavour
assignments
in the lattice simulations of the mesonic screening lengths that we have
reffered to. These were carried out using staggered fermions, and
in this scheme the space-time and flavour symmetries are intermingled
on the lattice and are retained only in the continuum limit.
The pion and sigma quoted originally by Detar and Kogut are related to
an  exact $U(1)\otimes U(1)$ symmetry of the free lattice fermion theory.
The generators corresponding to sigma and pi are $1\otimes 1$ and
 $\gamma_5 \otimes \tau_5$ respectively, and  the near degeneracy of their
screening masses above $T_c$ is interpreted as a restoration of the
symmetry. However this may be subject to some doubt, since, as first
pointed out by  Shuryak \cite{shur1}, only the connected part in Fig. 2a
was simulated on the lattice. The disconnected part in Fig. 2b remains to
be calculated.

 Since there is no lattice measurements of any other member in
the pseudoscalar multiplet we still do not know whether the whole continuum
$SU(4)_R \otimes SU(4)_L$ is restored, or only the above mentioned subgroup.
There is also no information on whether or not the axial $U_A(1)$ symmetry (not
to be confused with the lattice symmetry mentioned above) is restored.
This question, which is directly related to the strength of the anomaly at
finite temperature is important in order  to  correctly identify the
 symmetry group of the high temperature phase, and is thus crucial for
arguments of the type recently advanced by
Rajagopal and Wilczek \cite{raja2}. Clearly it would be very useful to
perform a direct finite temperature lattice calculation
 of the $U_A(1)$-types of order parameters such as
\be
{\rm det}\left (\psi^+ \half (1\pm\gamma_5) \psi\right ) \ \ \ \ \ . \nonumber
\ee
as originally suggested by t'Hooft \cite{thoo2}.

\vskip 3mm\noi
{\large\bf 3. Baryonic Correlators }
\renewcommand{\theequation}{3.\arabic{equation}}
\setcounter{equation}{0}

The DeTar correlator in the baryonic channels can be calculated completely
analogously to the mesonic ones. The currents appropriate for the nucleon and
the delta isobar, omitting color, flavour and space indices, are
\be
J^N &=& \left(\hpsi^T \hat C\ega_5\hpsi\right)\hpsi +\kappa
      \left(\hpsi^T \hat C\hpsi\right)\ega_5\hpsi \\
J^{\Delta}_{\mu} &=&
        \left(\hpsi^T \hat C\ega_{\mu}\hpsi\right)\hpsi
\ee
where $\kappa$ is an arbitrary mixing factor and $\hat C=\ega_2\ega_4$
the charge conjugation matrix. The existence of the two nucleon currents
(hence $\kappa$) was first noted by Ioffe \cite{ioff1}.
To use the dimensional reduction method we should transcribe (3.1) and (3.2) in
terms of the fields $\phi_\pm$, as shown in Appendix B.
Using these currents, the asymptotic form of the correlator (1.1) gives
the screening mass $m_\alpha$ in the baryonic channels.
To leading order $m_\alpha = 3M + E_{\alpha}$ is just
the mass of three heavy quarks in a color singlet configuration,
interacting via
 Coulomb forces. In the ground state, the three
quarks are in a symmetric spatial configuration, in which case (2.8)
simplifies to
\be
H_{ss} = \frac{e^2}{2M^2} \,\,(S_3^2 +\frac 34) \,\,
\delta^2(\vec r_{12})\ \ \ \ \ ,
\ee
where $S_3$ again is the total spin along
the 3-direction and $\vec r_{12}$ the relative two body separation in the three
body system. This is to be compared with the spin-spin interaction (2.9) in the
mesonic channel.

The ground state baryon wavefunctions are antisymmetric in color and symmetric
in spin, flavour and space, and are solution to the Faddeev
equations.  For simplicity, however, we will only make
variational estimates. In the center of mass frame we use the following
Gaussian variational ans{\"a}tze for the ground state wavefunction (we also
give results for high $T$ QCD in 2+1 dimensions\footnote{
As discussed in I, this theory, while
of no direct physical interest is easier to handle theoretically and perfectly
amenable to lattice calculations.}),
\begin{eqnarray}
QCD_{2+1}:&
&\psi (\rho, \eta ) =\sqrt{\frac{8 \alpha}{\pi}} e^{-\alpha (\rho^2 +\eta^2)}
\ \ \ \ \ , \\
QCD_{3+1}:&
&\psi (\rho, \eta ) =\frac {2 \alpha}{\pi} e^{-\alpha (\rho^2 +\eta^2)} \ \ \
\ \ ,
\end{eqnarray}
where $\vec\rho = \frac 1{\sqrt 2} (\vec r_1-\vec r_2)$ and
$\vec \eta = \sqrt{\frac 23}  (\vec r_1 + \vec r_2 - 2\vec r_3 )$.
%are the Jacobi coordinates. $\rho$ measures the relative
%separation between two quarks (here 1,2)  and $\eta$ the relative
%separation of the third quark (here 3) from the center of  mass.
Minimizing the expectation value of (2.7) in (3.4) and (3.5)
with respect to $\alpha$ yields for $QCD_{2+1}$,
$\alpha = \frac{1}{4} \left(\frac{9 e^4 M^2}{\pi}\right)^{\frac{1}{3}}$ and for
$QCD_{3+1}$, $\alpha = \frac{9 M e^2}{16\pi}$. The corresponding baryonic
screening lengths are,
\begin{eqnarray}
QCD_{2+1}:&
& m_\alpha = 3M + \frac{9}{4} \left(\frac{3 e^2}{M \pi} \right)^{\frac{1}{3}}
\ \ \ \ \ , \\
QCD_{3+1}:&
& m_\alpha = 3M + \frac {9e^2}{8\pi}
\left( 2- {\bf C} -{\rm ln }(\frac{e^2}{16\pi M})\right) \ \ \ \ \ ,
\end{eqnarray}
where ${\bf C} \approx 0.577$ is  Euler's constant. The spatial distribution
of the baryonic correlators is Gaussian in both $\vec\rho$ and $\vec\eta$, and
have widths $\sim 4\pi/e\sqrt M$. Another simple
alternative to a full solution to the Faddeev equations is to use a
quark-diquark model as shown in Appendix C.

Like in the mesonic case, the degeneracy of the nucleon and delta screening
masses will be  lifted by spin dependent effects. At very high temperatures we
expect the splitting to be due to the spin-spin interaction (3.3). This
gives an upward shift to $both$ the nucleon and the  delta.
The removal of the degeneracy is, however, only
partial  as the $\Delta (3/2, \pm 1/2)$ states remain degenerate with
the $N(1/2,\pm 1/2)$ states. We get
\be
m^{\pm 1/2}_N - m_{\alpha} = m_{\Delta}^{\pm 1/2} -m_{\alpha} =
\frac 13 \left( m_{\Delta}^{\pm 3/2} - m_{\alpha} \right)
\ee
and the splitting is $\sim e^2/4M^2$. As already mentioned,
the different polarization's of the vector and axial vector mesons, are split
so
the $j = \pm 1$ states are pushed up, while the $j=0$ state is
unaffected along with the pion and its scalar partner the $\sigma$.
At very high temperature where this effect should be dominant, we thus predict
\be
m_{\Delta}^{\pm 3/2} - m_{\Delta}^{\pm 1/2} =
m_{\Delta}^{\pm 3/2} - m_N^{\pm 1/2}
\ee
of the same order of magnitude and $direction$ as the meson splitting
\be
m_{\rho}^{\pm 1} - m_{\rho}^0 = m_{\rho}^{\pm 1} -m_{\pi}
\ee
Since chiral symmetry is manifest at high temperature, these relations
also hold for the chiral partners.

\vskip 3mm\noi
{\large\bf 4. Gluonic Correlators }
\renewcommand{\theequation}{4.\arabic{equation}}
\setcounter{equation}{0}

In this section we discuss correlators  of local operators containing
gluon fields. Color singlets are easily constructed from the field strength
tensor $G_{\mu\nu}$, and its covariant derivatives. We shall only consider
the operators $E^2 $ and $B^2$, which are the lowest dimensional operators that
couple to the two gluon channel, and the trace of the Polyakov loop,

\be
L(x) ={\rm P}e^{ig\int_0^{1/T}d\tau A_0 (\tau, \vec x)}\ \ \ \ \ , \nonumber
\ee
where P denotes path-ordering,
which has been extensively studied by  lattice calculations.

At finite temperature, $E^2 $ and $B^2$ are distinct operators since Lorentz
invariance is broken. In the dimensional reduction scheme, this is manifest
by the presence of both a (2+1) dimensional vector gluon field, $A_\mu$ and a
scalar adjoint Higgs field, $H$, in the lagrangian (2.5). The operator $E^2$
couples to the two-particle channel $(HH)$ and $(A_\mu A_\mu)$
with strength 1 and $g^2$ respectively, while $B^2$ only couples to
$(A_\mu A_\mu)$, see fig. 1. Since $g \rightarrow 0$ at high $T$, we would
thus naively
expect the screening mass $2m_E$ in the $(E^2\, E^2)$ correlator and a
power-law
fall off in the $(B^2\,B^2)$ one. Of course we know that this is
oversimplified. As already discussed there are lots of evidence that the
dimensionally reduced theory is confining so it makes no sense to treat the
magnetic sector, \ie the $A_\mu$ field, in perturbation theory. We might hope,
however, that the $(HH)$ channel can be treated in the same way as the
$\overline q q$ one. In that case the above asymptotic (\ie $T \rightarrow
\infty$) prediction
\be
\langle E^2(x^i,x^3) E^2(0,0)\rangle \sim K_1 (m_{E^2} x^3)
\ee
where $m_{E^2} = 2m_E  = \frac 4 {\sqrt 3} g T$, and $K_1$ a modified
Bessel function,
should be good at least for some moderately large values of $x_3$. (For really
large distances there will always be an uncontrollable contamination by the
magnetic sector since the small coupling strength will be compensated for by a
larger exponential factor. This is not the case in 3d QCD at high temperature
where we expect the above result to hold with the substitution $K_1\rightarrow
K_0$). It is important to stress that if it were not for the correlations in
the transverse directions, (4.1) would decay as $K_1^2 (m_E x^3)$,
which has the same exponential fall-off as (4.1). It is the
{\em pre-exponent} factor that distinguishes between $bound$ and $free$
electric gluons
at asymptotic temperatures.

The binding energy is easily calculated using formulae similar to (2.7) and
(2.8)  modified for the particles having octet color charge and being scalars.
A simple test of the self consistency of the estimate (4.1) is that the radius
of the corresponding bound state is small enough for the magnetic effects to
be ignored. Using the same variational estimates as in the mesonic case yields
bound state radii $ \sim 1/g^{3/2}T$, which is in between the electric and
magnetic length scales $1/gT$ and  $1/g^2T$ respectively. We
consider it an open question whether the method based on dimensional reduction
is reliable in this case. We should also mention that $E^2$ also couples to
three and four particle channels, but these contributions are again suppressed
by exponential's and/or powers of $g$.

We already concluded that the operator $B^2$, is beyond the range of
applicability of our method. If, however, the main effect of the infrared
divergences in the magnetic sector is the generation of a magnetic mass $m_M
\sim g^2T$, we have the qualitative prediction that asymptotically the $B^2$
screening length is larger than the $E^2$, since the magnetic
mass is smaller than the electric.
It is clearly of interest to simulate both these correlators on the lattice.

We now turn to correlators of Polyakov loops (or Wilson lines),
\be
\langle {\rm Tr} {\rm L}^{\dag}(\vec x) {\rm Tr} {\rm L}(\vec 0)\rangle
\rightarrow N_c^2 e^{-V_{\bf 1}(T,x)/T}
\label{polya}
\ee
corresponding to inserting two static quarks in a heat bath of gluons.
$V_{\bf 1}$ in (\ref{polya}) is the singlet potential.
In the confined phase the singlet potential is believed to
behave as $\sigma (T) x$ for large $x= |\vec x|$, where $\sigma(T)$ is
the (temperature dependent) string tension.
In the deconfined high temperature phase the potential is usually believed to
be screened,
\be
V_{\bf 1} (T, \vec x) = 2F_Q (T) + V_{1,0} (T) e^{-2m_E  x} \ \ \ \ \ .
\label{scree}
\ee
The exponential fall off is again governed by the two (electric) gluon channel
just as for the $E^2$ correlator.  As pointed out by Nadkarni \cite{nadk1},
 also in
this case we expect contamination from the magnetic sector at large enough
distances. However, using the same arguments as for $E^2$ we suggest that the
screening length should be calculable in the dimensionally reduced theory as an
$(HH)$ bound state. Clearly, one would need very precise measurements
to test whether the non-relativistic bound state interpretation of the gluonic
correlators is correct. In particular it would be interesting to simulate
the above correlator on the lattice and compare the results with the
Wilson like loop at finite temperature.

Recently there have been lattice simulations aimed at probing the light
fermion distributions ($\overline{\psi}\psi$ and $\psi^{\dag}\psi$)
\cite{deta1}.
Specifically, it has been  observed that the fermion number distribution
around a heavy quark given by
\be
\langle {\rm Tr}{\rm L} (\vec 0) \psi^{\dag}\psi (\vec x)\rangle
\label{fermion}
\ee
is localized around the heavy source (with a weight of minus one) at
low temperature but spread out at high temperature. We expect that this spread
is related to the screened two-gluon (dipole) exchange with a range of
the order of $1/2m_E$ as discussed above.
The correlation function (4.4) is consistent with Gauss' law in the following
sense. In the presence of light fermions the sampled gauge configurations on
the lattice have an excess or a deficiency of color charge (in the sense of
a grand canonical colored ensemble). When introducing a heavy triplet source,
the configurations with an anti-triplet or a sextet of color are sampled out
so that on the average Gauss law is enforced. The sampled configurations
together with the heavy source carry zero triality.\footnote{
We thank Carleton DeTar for a discussion on this point.}

\vskip 3mm\noi
{\large\bf 5. Nonasymptotic Effects }
\renewcommand{\theequation}{5.\arabic{equation}}
\setcounter{equation}{0}

As already mentioned, our estimate of the fine structure of the screening mass
spectrum is only reliable at asymptotically high temperature. Close to the
transition region, where the available lattice data are collected, many
other effects are possible. Examples are Higgs interactions,
effects  induced by a dilute gas of monopoles,  a residual rarefied gas of
instantons and antiinstantons, nonvanishing magnetic condensates...
Such effects are unfortunately much harder  to  calculate in a controlled
manner, but we shall nevertheless attempt some estimates.

At zero temperature, the presence of gluon condensates modifies the
perturbative spin-spin interaction and gives a contribution to the spin-spin
splitting $\sim \langle B^2\rangle$ \cite{leut1},\cite{volo1}.
These effects will presumably persist even at finite temperature, but since we
lack information about the condensates they are hard to estimate. However, the
symmetry structure of these terms would be the same as for the perturbative
spin-spin interaction (2.8).

\newcommand\phib {\overline\phi}
The effects due to Higgs exchange is rather straightforward to estimate.
First we consider the $m=0$ case where the Higgs couple only to the scalar
density $\phib \phi$. To
lowest order, we just have to consider the exchange of a scalar octet particle
with mass $m_E$. The coupling is only to the color charge, so this will not
cause any level splitting. Rather than trying to solve the Schr{\"o}dinger
equation in a potential including the Yukawa potential from the Higgs
exchange, we consider the latter as a perturbation. If we furthermore neglect
the propagation of the Higgs particle, \ie make the local approximation
$1/m_E^2$ for the propagator, we get the following local potential,
\be
V_{Higgs} = \frac {e^2} {m_E^2} \delta^2(\vec r)
= \frac 1 T  \delta^2(\vec r) \ \ \ \ \ ,
\ee
where the last equality follows for two light quarks. This interaction will
shift the meson screening masses by
\be
\Delta m = \frac 2 T |\psi(0)|^2
\ee
and comparing with the $\pi - \rho$ mass-shift due to the spin-spin splitting
\be
m^{\pm}_{\rho} - m_\pi = \frac 2 3 \frac {g^2} {\pi^2} \frac 1 T |\psi(0)|^2\ \
\ \ \ ,
\ee
we see that the Higgs effect is an order of magnitude larger at $\alpha_s =
0.1$ . Similar estimates of the level-shifts due to Higgs exchange can be made
for the baryon and gluon correlators.

In order to split the rho from
the pi without splitting the polarization components of the rho, we need a
coupling to $\phib\gamma_3 \phi$. A possible, and even plausible, scenario is
that the effective lagrangian for the quarks includes an effective four fermi
interaction of the Nambu Jona-Lassinio type. Below the transition temperature
such an interaction is known to induce a  spontaneous breaking of chiral
symmetry
(at the mean field level) and generate a
constituent quark mass. It is possible that the same interaction could be
responsible for the large observed splitting of the pi and rho screening
masses. Neglecting flavour (which is anyhow not correctly represented in the
lattice calculations) we can take
\be
{\cal L}_{NJL} = \lambda \left[ (\phib\gamma_5\phi) -
 (\phib \gamma_3 \phi)^2 \right]
\ee
which is chirally invariant. The corresponding spin-spin interaction between a
quark and an antiquark is of the form $\sim \lambda S^i_1 S^i_2$ which is
appropriate for pushing down the pi and leaving the rho untouched. The
interaction is however too singular to be treated in perturbation theory. A
correct treatment, which would amount to resolving the Schr{\"o}dinger equation
with a pointlike interaction, will not be attempted here.

It is also instructive to consider the limit $M\gg T$ (heavy quark limit).
In this case the
coupling to the Higgs is entirely due to the density $i\phib\gamma_3\phi$,
which reduces to $1/2M
\chi^\dagger(\epsilon^{ij} q_i\sigma_j)\chi$ in the nonrelativistic limit,
where $\chi $ is a two-spinor and
$\vec q$ the momentum transfer. The resulting spin-spin interaction due to
the Higgs is of the form
 \be
H_{ss,H} =\sum_{i<j} \frac {e^2}{M^2}(S_1^iS_1^j +S_2^iS_2^j)
\left(\delta^2 (\vec{r}_i -\vec{r}_j) -\frac{m_E^2}{2\pi} K_0 (m_E |\vec{r}_i
-\vec{r}_j|)\right)\ \ \ .
\ee
The local part in (6.5) combines with the local part of (2.8)
to give

\be
H_{ss} = \sum_{i<j} \frac {e^2}{4M^2} \left( 1+4S^i\cdot S^j\right)
\delta^2 (\vec{r}_i-\vec{r}_j)
\ee
So, as expected, for heavy quarks, we recover the conventional
polarization independent
spin-spin interaction. This result is also supported by the lattice results.
For mesonic systems we have
$H_{ss} =(2S^2-1)e^2/4M^2$, while for baryonic systems we have
$H_{ss} =(2S^2-3/2)e^2/8M^2$. This spin-spin interaction pushes the
rho up, the
 pion
down, leaves the nucleon unchanged and pushes the delta up. Specifically,
$m_{\rho}-m_{\pi} =e^2/M^2$ and $m_{\Delta}-m_N =3e^2/4M^2$. In this regime
we expect

\be
m_{\Delta}-m_N = \frac 34 (m_{\rho}-m_{\pi}) \ \ \ \ \  .
\ee
It is not ruled out, that the lattice results for the sources quoted
are carried out with still large quark masses, in which case the present
discussion is pertinent.

Finally we give a  rough estimate of the effect of instantons assuming
that as chiral symmetry is restored,  the
instanton-antiinstanton medium gets rarefied (single particles, molecules,
small clusters). The effects would be to push down the nucleon
but leave the delta unaffected. Since the instanton contribution
is strongest in the quark-antiquark or quark-quark spin-isospin zero channel,
we would expect a delta-nucleon splitting in the screening lengths
in comparison to the $\rho-\pi$ splitting of the order of
\be
m_{\Delta} -m_N\sim \frac 3 2 (m_{\rho}-m_{\pi})\ \ \ \ \ .
\ee
Above the transition temperature this splitting is entirely due to
the nonvanishing of the strange quark mass. The splitting vanishes
exponentially at asymptotic temperatures. This splitting is comparable
in magnitude and direction to the spin-spin splitting discussed above
for heavy quarks.
\footnote{We recall that the splitting is proportional to the instanton density
which is proportional to $e^{-4/3 \pi^2 \rho^2 T^2}$ where $\rho$ is the
instanton size. For temperatures of the order of $T_c$ the density drops
by an order of magnitude \cite{pisa3},\cite{gros2}.  }

\vskip 3mm\noi
{\large\bf 6. Discussion }
\renewcommand{\theequation}{7.\arabic{equation}}
\setcounter{equation}{0}

We have presented simple estimates of the baryonic screening
lengths and spatial extension using arguments on dimensional reduction.
These arguments are straightforward extensions of our previous arguments
for mesons.
We have also provided some estimates for the gluonic correlation
functions, and suggested new correlation functions to be evaluated
on the lattice.

While the dimensional reduction scheme is only valid at asymptotic
temperatures, we have also used it to  estimate  nonasymptotic
effects caused by both perturbative gluons (spin-spin interaction) and
nonperturbative gluons (instantons). It would be interesting to see
how these predictions compare with lattice simulations for the hadronic
correlation functions. We stress that our results are only suggestive
since the calculations may be sensitive to the magnetic length scale.
If, however, future lattice calculations will confirm, or at least
support, our calculations, our calculations of nonasymptotic effects
could be much improved. Since the problem is essentially that of bound
states in 2+1 dimensional QCD, we can imagine to use many of the
phenomenological
methods, like sum rules, bags {\em etc.},  that have been developed to
calculate
 bound states in 3+1 dimensions. It remains to be seen whether such a
''high $T$ bound state phenomenology'' will become feasible.

There are several places for improvement in our calculations. Most
noteworthy: if our estimates  of  sizes and binding energy are correct, they
are midway between the electric and magnetic scale. Thus, it is important
to do a more complete calculation including hard thermal loops to  better
 assess  these  effects. Since all the correlators discussed
in this work are gauge invariant, the tricky issue of gauge dependence
related to the subdivision into hard ($g^2T$) and soft ($gT$) does not
arise. This exercise is worth pursuing in both  four and three
dimensional QCD. Again, lattice measurements are crucial to settle these
questions.

\vskip 5mm\noi
{\large\bf  Acknowledgements}

We thank Carleton DeTar and Bengt Petterson for explaining their
staggered lattice fermion calculations to us.

%\newpage
\vskip 5mm\noi
{\large\bf Appendix A. Two-component formulation }
\renewcommand{\theequation}{A.\arabic{equation}}
\setcounter{equation}{0}
\newcommand\psib {\overline\psi}

In this appendix we shall show how to formulate the dimensionally
reduced theory
in a two-component formalism. We again start from the Lagrangian (2.1)
(for simplicity we only consider the lowest positive node), but we
now rotate to Minkowski space by the transformations
\be
\hpsid \rightarrow i\overline\psi = i\psi^\dagger\gamma^0
\ee
where $\gamma^0 = -\ega_3$. We also introduce the transverse Minkowski gamma
matrices by $\ega_i = i\gamma^i$ to get
\be
{\cal L} = \psib (i\gamma^0\partial_0 + i\gamma^i\partial_i - i\ega^0 M)\psi
\ee
After introducing the two two-spinors $\psi_1$ and $\psi_2$ via
\be
\psi = \left( \begin{array}{c} {\psi_1} \\ {\psi_2} \end{array} \right)
\ee
the Lagrangian takes the form
\be
{\cal L} = \psib_1 i\gamma_3^i\partial_i \psi_1 +
 \psib_2 i\gamma_3^i\partial_i \psi_2 - M (\psib_1\psi_2 + \psib_2\psi_1)
\ee
where the 3-d gamma matrices are defined by $(\gamma_3^0,
\gamma_3^1,\gamma_3^2) = (\sigma^3,i\sigma^1,i\sigma^2)$, and $\psib_i =
\psi^\dagger_i\gamma_3^0$. We leave it as an exercise to put in the gluons and
a mass for the quark.  (A.4) has the form of a pair of standard
2+1 dimensional Dirac Lagrangian's, but with an off-diagonal mass term. This
means that in order to obtain a propagator one must diagonalize a
four-dimensional matrix anyway, and it is more convenient to use the
representation given in section 2.

\vskip 3mm\noi
{\large\bf Appendix B. Quantum numbers of baryonic currents }
\renewcommand{\theequation}{B.\arabic{equation}}
\setcounter{equation}{0}

There are two inequivalent local sources for spin $\frac 12$ baryons
of opposite parity

\be
B_1^+ &=& \left(\hpsi^T \hat C\hat\gamma_5\hpsi\right)\hpsi \nonumber\\
B_2^+ &=& \left(\hpsi^T \hat C\hpsi\right)\hat\gamma_5\hpsi
\ee
\be
B_1^- &=& \left(\hpsi^T \hat C \hpsi\right)\hpsi \nonumber\\
B_2^- &=& \left(\hpsi^T \hat C \hat\gamma_5\hpsi\right)\hat\gamma_5\hpsi
\ee
For convenience, we will throughout omit reference to
color, flavour and space indices.
Under parity ($\hpsi\rightarrow \ega_4\hpsi$) : $B^{\pm}\rightarrow
\pm \ega_4 B^{\pm}$. $B^+$ and $B^-$ refer to the octet ${\frac 12}^+$
and ${\frac 12}^-$ baryons respectively. The nucleon is an arbitrary linear
combination of $B_{1,2}^+$, $i.e.$ $J^N = B_1^+ +\kappa B_2^+$.

In the
chiral basis the combinations $B_1^{\pm} \pm B_2^{\pm}$ are more appropriate
to use. In this basis the nucleon can be viewed either as $(RRR)-(LLL)$ or
$(RRL)-(LLR)$ or any linear  combination of the two. The odd parity partners
of the nucleon are $(RRR)+(LLL)$ or $(RRL)+(LLR)$ or again any
linear combination
of the two. This representation is suited to the two component formulation
discussed in Appendix A.

To exhibit the structure of the currents in the four-component formulation
discussed in the text, we first derive the reduced forms for the diquark
constituents. Specifically

\be
\beta\hpsi^T\hat C\ega_5\hpsi &=&
-\hphi_-^T\ega_1\ega_3 U_-^2\hphi_- \,e^{-i2\pi T \hat x_4}
-\hphi_+^T\ega_1\ega_3 U_+^2\hphi_+ \,e^{i2\pi T \hat x_4}\\
&-&\hphi_-^T\ega_1\ega_3\hphi_+
  -\hphi_+^T\ega_1\ega_3\hphi_-  \nonumber \\
\beta\hpsi^T\hat C\hpsi &=&
\hphi_-^T\ega_2\ega_4 \hphi_-  \,e^{-i2\pi T \hat x_4}
+\hphi_+^T\ega_2\ega_4 \hphi_+  \,e^{i2\pi T \hat x_4}\\
&+&\hphi_-^T\ega_2\ega_4 U_+^2\hphi_+
  +\hphi_+^T\ega_2\ega_4 U_-^2\hphi_-   \nonumber
\ee
The unaveraged sources for the baryons follow from these diquark structures
and the reduced fermionic fields (2.3). The
resulting expressions involve terms with the $\tau$-dependence
$e^{\pm i3\pi T\tau}$ and
$e^{\pm i\pi T\tau}$. Static sources can be constructed by averaging over
$\pi T$ or $3\pi T$. We choose the latter and define
 \be
\overline{B}^{\pm}_{1,2} =
T\,\int_0^{\frac 1T} d\hat x_4\, {\rm cos}(3\pi T\hat x_4 ) B_{1,2}^{\pm}
\label{static}
\ee
the lattice calculations have been carried out by averaging over the lowest
Matsubara frequency $\pi T$. Using (\ref{static}), the reduced
diquark sources (B.3) and (B.4),
and the fermionic fields (2.3), we obtain
 \be
2\beta ^{\frac{3}{2}}\overline{B}_1^+ &=&
-i\left( \hphi_-^T\ega_1\hphi_-\right)V U_-\hphi_- +
 i\left( \hphi_+^T\ega_1\hphi_+\right)V U_+\hphi_+\nonumber\\
2\beta ^{\frac{3}{2}}\overline{B}_2^+ &=&
\left( \hphi_-^T\ega_2\ega_4\hphi_-\right) \ega_5 V U_-\hphi_- +
\left( \hphi_+^T\ega_2\ega_4\hphi_+\right) \ega_5 V U_+\hphi_+
\ee
\be
2\beta ^{\frac{3}{2}}\overline{B}_1^- &=&
\left( \hphi_-^T\ega_2\ega_4\hphi_-\right) V U_-\phi_- +
\left( \hphi_+^T\ega_2\ega_4\hphi_+\right) V U_+\hphi_+\nonumber\\
2\beta ^{\frac{3}{2}}\overline{B}_2^+ &=&
-i\left( \hphi_-^T\ega_1\hphi_-\right) \ega_5 V U_-\hphi_- +
 i\left( \hphi_+^T\ega_1\hphi_+\right) \ega_5 V U_+\hphi_+
\ee
Our static sources do not mix the $\phi_+$ and $\phi_-$ fields while
the lattice sources averaged over $\cos (\pi T\hat x_4)$ do.
It would be interesting to find out
how (if at all) this affects the structure of the resulting correlation
functions.

In the case of the isobar current, there is a unique local current
$(\hpsi^T\hat C\ega_{\mu}\hpsi)\hpsi$. Its form in the four component
formalism can be obtained without difficulty using techniques similar
to the ones  above, and will not be given here. We note, however,
that the isobar carries a polarization. The isobar diquark field
is just $\hpsi^T\hat C\ega_{\mu}\hpsi$. This means that the sources
for the transverse and longitudinal fields are distinct at high temperature
and are subject to the same remarks developed in the text for the vector
meson fields. In particular, we expect the subleading spin-spin interactions
to affect the longitudinal and transverse parts of the isobar differently.

\vskip 3mm\noi
{\large\bf Appendix C. Quark-diquark model for baryons }
\renewcommand{\theequation}{C.\arabic{equation}}
\setcounter{equation}{0}

If we were to use the quark-diquark picture for baryons than the
screening lengths and wavefunctions are readily obtained from the
mesonic screening lengths and wavefunctions. In the case of a free
diquark of mass $M_* = 2M$ with antitriplet charge, a variational
estimation shows that the screening mass is,
\begin{eqnarray}
QCD_{2+1}:&
&m_{\alpha} = 3M + \frac{3}{2}
\left( \frac{3 e^4}{16 \pi M} \right)^{\frac{1}{3}} \\
QCD_{3+1}:&
&m_{\alpha} = 3M + \frac {e^2}{8\pi}\left( 2-{\bf C}-
                {\rm ln}(\frac {e^2}{12M\pi} )\right)
\end{eqnarray}
In the antitriple channel, the diquark mass is,
\begin{eqnarray}
QCD_{2+1}:&
&M_* = 2M +\frac{3}{2} \left( \frac{e^4}{4 \pi M} \right)^{\frac{1}{3}}
\\
QCD_{3+1}:&
&M_* = 2M +\frac {e^2}{8\pi}\left( 2 -{\bf C}-
                 {\rm ln} (\frac{e^2}{8\pi M})\right)
\end{eqnarray}
The baryonic screening lengths in this case are,
\begin{eqnarray}
QCD_{2+1}:&
&m_{\alpha} = M +M_*
+\frac{3}{2}
\left( \frac{e^4}{8 \pi} \frac{(M+M_*)}{M M_*} \right)^{\frac{1}{3}} \\
QCD_{3+1}:&
&m_{\alpha} = M +M_*
+\frac {e^2}{8\pi}\left( 2 -{\bf C} -
                    {\rm ln} (\frac{e^2}{2\pi(M+M_*)}) \right)
\end{eqnarray}
In the diquark picture, the baryonic wavefunctions are similar to the
mesonic wavefunctions.

%\newpage
%\bibliography{[hansson.paper.bib]field}

\begin{thebibliography}{10}

\bibitem{gava1}
{R.V. Gavai,} et~al,
\newblock {\it Phys. Lett.}, {\bf B241} (1990)567.

\bibitem{kars1}
F. Karsch,
\newblock {\it Nucl. Phys. (supp)}, {\bf 9B} (1989)357.

\bibitem{gott1}
{S. Gottlieb,} et~al,
\newblock {\it Phys. Rev. Lett.}, {\bf 59} (1987)2247.

\bibitem{bern1}
{C. Bernard,} et~al,
\newblock {\it Phys. Rev. Lett.}, {\bf 68} (1992)2125.

\bibitem{mano1}
E. Manousakis and J. Polonyi,
\newblock {\it Phys. Rev. Lett.}, {\bf 58} (1987)847.

\bibitem{kark1}
{L. Karkkainen,} et~al,
\newblock {\it Phys. Lett.}, {\bf B312} (1993)173.

\bibitem{fing1}
{J. Fingberg,}~U. Heller and F. Karsch,
\newblock {\it Nucl. Phys.}, {\bf B392} (1993)493.

\bibitem{hans5}
T.H. Hansson and I. Zahed,
\newblock {\it Nucl. Phys.}, {\bf B374} (1992)117.

\bibitem{prak1}
M. Prakash and I. Zahed,
\newblock {\it Phys. Rev. Lett.}, {\bf 69} (1992)3282.

\bibitem{zahe1}
I. Zahed,
\newblock {\it Hot QCD},
\newblock Preprint, SUNY-NTG-93-42, Act. Phys. Pol. B (1994) in print.

\bibitem{deta2}
C. DeTar and J.B. Kogut,
\newblock {\it Phys. Rev.}, {\bf D36} (1987)2828.

\bibitem{brow2}
{G.E. Brown,} et~al,
\newblock {\it Phys. Rev.}, {\bf D45} (1992)3169.

\bibitem{shur1}
E. Shuryak,
\newblock {\it Private Communication}.

\bibitem{raja2}
K. Rajagopal and F. Wilczek,
\newblock {\it Nucl. Phys.}, {\bf B404} (1993)577.

\bibitem{thoo2}
G. t'Hooft,
\newblock {\it Phys. Rev.}, {\bf D14} (1976)3532.

\bibitem{ioff1}
B. Ioffe,
\newblock {\it Nucl. Phys.}, {\bf B188} (1981)317.

\bibitem{nadk1}
S. Nadkarni,
\newblock Debye screening and nonperturbativity in {Hot QCD},
\newblock In {\it Brookhaven, Lattice gauge theory}, 1986.

\bibitem{deta1}
C. DeTar,
\newblock {\it Nucl. Phys. (supp)}, {\bf 30B} (1993)66.

\bibitem{leut1}
H. Leutwyler,
\newblock {\it Phys. Lett.}, {\bf B98} (1981)447.

\bibitem{volo1}
M.~B. Voloshin,
\newblock {\it Nucl. Phys.}, {\bf B154} (1979)365.

\bibitem{pisa3}
R.D. Pisarski and L.G. Yaffe,
\newblock {\it Phys. Lett.}, {\bf B97} (1980)110.

\bibitem{gros2}
{D.J. Gross,}~R.D. Pisarski and L.G. Yaffe,
\newblock {\it Rev. Mod. Phys.}, {\bf 53} (1981)43.

\end{thebibliography}

\newpage
{\bf Figure Captions}
\vskip 5cm
\noindent Fig. 1 : Typical couplings to $E^2$ (a) and $B^2$ (b) to the electric
                   and magnetic gluons.

\vskip 5cm
\noindent Fig. 2 : (a) Direct contribution to the sigma-sigma correlator;
                   (b) exchange contribution to the sigma-sigma correlator.

\end{document}